\documentclass[a4paper,11pt]{article}
\usepackage{amsmath, amsthm, amsfonts, amssymb}
\usepackage[svgnames]{xcolor}
\colorlet{color1}{NavyBlue}
\usepackage[colorlinks=true,allcolors=color1]{hyperref}
\usepackage{physics}
\usepackage{dsfont}
\usepackage{graphicx}
\usepackage{cleveref} %use \cref{label}
\usepackage{xfrac}
\usepackage{arydshln}
\usepackage{subcaption}
\usepackage{orcidlink}

\allowdisplaybreaks

\title{Semiclassical evolution of a dynamically formed spherical black hole with an inner horizon}

\date{}
\author{Valentin Boyanov~\orcidlink{0000-0002-2458-7200}, David Hilditch~\orcidlink{0000-0001-9960-5293}, Artur Semião~\orcidlink{0000-0002-2399-2220} \vspace*{.2cm}\\ \textit{CENTRA, Departamento de F\'{\i}sica, Instituto Superior T\'ecnico -- IST,}\\ \textit{Universidade de Lisboa -- UL, Avenida Rovisco Pais 1, 1049 Lisboa, Portugal}}

\begin{document}
\maketitle

\begin{abstract}
  In this work we obtain a numerical self-consistent spherical
  solution of the semiclassical Einstein equations representing the
  evaporation of a trapped region which initially has both an outer
  and an inner horizon. The classical matter source used is a static
  electromagnetic field, allowing for an approximately
  Reissner-Nordström black hole as the initial configuration, where
  the charge sets the initial scale of the inner horizon. The
  semiclassical contribution is that of a quantum scalar field in the
  ``in" vacuum state of gravitational collapse, as encoded by the
  renormalised stress-energy tensor in the spherical Polyakov
  approximation. We analyse the rate of shrinking of the trapped
  region, both from Hawking evaporation of the outer apparent horizon,
  as well as from an outward motion of the inner horizon. We also
  observe that a long-lived anti-trapped region forms below the inner
  horizon and slowly expands outward. A black-to-white-hole transition
  is thus obtained from purely semiclassical dynamics.
\end{abstract}

\section{Introduction}

Astrophysical black holes (BHs) are among the most analysed objects in
theoretical gravitational physics. As we gain the ability to observe
the characteristics of the spacetime these objects generate with more
and more accuracy~\cite{LIGO,EHT,LISA,ngEHT}, it is becoming
increasingly crucial to understand their fundamental nature. Classical
General Relativity (GR) offers a consistent description of their
formation and evolution as seen from the outside, given that horizons
always seem to cover up the singular region, where the theory
eventually breaks down. However, the very fact that such a breakdown
region exists, combined with our knowledge of the fundamentally
quantum nature of matter, makes it clear that this is not the whole
picture.

A first approach toward obtaining modifications to classical
gravitational dynamics can be obtained just by attempting to describe
the propagation of quantum fields on curved spacetime backgrounds, and
taking into account the backreaction of their averaged stress-energy
content on the evolution of the
geometry~\cite{BirrellDavies}. Applying this semiclassical approach to
geometries of BH formation, it was found that corrections to the
classical picture can significantly alter the causal structure of the
spacetime. Particularly, after horizon formation, an initially empty
quantum vacuum state acquires an energy content in the form of a
positive outgoing flux, which moves away from the BH, compensated by a
negative ingoing flux, which slowly depletes the mass and reduces the
size of the BH~\cite{Hawking1975,DFU1976}. Thus, despite the spacetime
curvature around the horizon typically being far from the Planck
scale, there is nonetheless an important cumulative effect which
changes the horizon dynamics, and consequently the global causal
structure.

Hawking evaporation is an example around which two important changes
to our understanding of BHs were brought about. On the one hand, the
fact that leading-order quantum corrections to classical gravitational
dynamics need not be local in curvature, i.e. significant corrections
need not be restricted to the vicinity of classical singularities,
particularly in the presence of horizons. On the other hand, the fact
that the classical modelling of the interior of these objects is no
longer sufficient, as whatever the classically singular region may be,
it would eventually be brought into causal contact with the outside
universe.\footnote{Except perhaps if the final stage of their
  evolution is an extremal
  configuration~\cite{DiFilippo2024}. However, as we will discuss
  below, based on our present results this possibility seems unlikely
  within our class of initial data.}

These two points are once again brought together when analysing
another region of black hole geometries: the \textit{inner} apparent
horizon. In the absence of matter, BHs are generically expected to be
described by the Kerr-Newman metric~\cite{Wald1984GR}, in which there
is always an inner horizon (except in the measure zero subset of
Schwarzchild BHs). In the maximal analytical extension of these vacuum
spacetimes, the inner horizon coincides with a Cauchy horizon of the
spacetime, while in gravitational collapse there is generally a
dynamical inner apparent horizon, which only asymptotes to a Cauchy
horizon in the absence of further dynamics (see fig.~\ref{f1}).

Semiclassical analyses of black holes with inner
horizons~\cite{Birrell1978, BalbinotPoisson93, Zilberman2019,
  Zilberman2022, Hollands2019, Hollands2020, McMaken2024} have thus
far mainly been carried out on the fixed backgrounds of maximal
extensions, where the finite lifetime of the trapped region is not
taken into account, and where the inner horizon is already a Cauchy
horizon, a surface known for its divergent blueshift effect on field
perturbations~\cite{Simpson1973}. These works then conclude that if
the formation of a Cauchy horizon were approached, then backreaction
from the Renormalised Stress-Energy Tensor (RSET) of quantum fields
would begin to dominate the evolution and would likely lead to the
formation of a (strong) curvature singularity. This, however, is not
very informative of what semiclassical dynamics looks like in a
geometry where an inner horizon has formed dynamically from
gravitational collapse, since a semiclassical Cauchy horizon formation
is highly doubtful, even if one only considers the well-established
Hawking evaporation process.

To address this issue, one would have to compute the RSET of a field
in a dynamical collapse background, which would pose additional
technical challenges~\cite{Anderson2020}, which, to our knowledge,
have not yet been fully resolved and implemented (although
see~\cite{Ori2025} for a recent result). As a first step toward
gaining insight as to what the full backreacted evolution of such
trapped regions would be, one of the present authors and collaborators
performed analytical calculations using the Polyakov
approximation~\cite{Fabbri2005} to the RSET of a massless scalar field
in a simple spherically-symmetric shell collapse
model~\cite{Barcelo2021}. It was found that, much like the outer
horizon has a tendency to shrink and reduce the size of the trapped
region from the outside, the inner horizon has a tendency to expand
outwards, reducing the size of the trapped region from the inside. It
was then conjectured that if the non-linear backreaction which takes
place for classical perturbations (leading to the mass inflation
instability~\cite{Poisson1989,Ori1991,Hod1998,Dafermos2017}) is also
triggered with the semiclassical source, and if at late times it is
the semiclassical source which dominates, then the trapped region may
end up disappearing completely due to a rapid outward inflation of the
inner horizon, on timescales much shorter than the Hawking evaporation
time~\cite{Barcelo2021,Barcelo2022}.

In the present work, we again use the Polyakov approximation for the
RSET and construct an ``in" vacuum state with a collapsing shell which
leads to the formation of a Reissner-Nordström black hole. We then
numerically evolve the full semiclassical Einstein equations in
spherical symmetry, with a classical source in the form of the
electromagnetic field generated by the central charge, along with the
Polyakov RSET of a scalar field on the semiclassical side, analogously
to the Schwarzschild geometry case studied in~\cite{Parentani1994}. In
the evolution we find both the expected inward tendency of the outer
horizon, as well as the outward tendency of the inner horizon, both of
which continue until the trapped region disappears
completely. However, in this particular setup an exponential
instability of the inner apparent horizon does not seem to manifest;
rather, its outward radial motion is only mildly faster than the
inward motion of the outer horizon, the latter of which can be
attributed to Hawking evaporation. Indeed the total evaporation time
of the trapped region appears compatible with the cubic law of Hawking
evaporation~\cite{Hawking1975} (for different cases where the charge
to mass ratio is kept the same, and the charge is kept constant
throughout each evolution).

Aside from the complete evaporation of the trapped region, the
numerical evolution of this model reveals an additional feature: the
appearance of an anti-trapped region below the trapped one, with an
outer marginally anti-trapped surface which slowly expands radially
outward until it approximately reaches the size of the last trapped
surface. This anti-trapped region continues to be present after the
full evaporation of the trapped region, making this model effectively
a black-to-white-hole transition. The end state of the evolution will
be the subject of future investigation. While our setup is similar to
the two dimensional dilaton gravity semiclassical model presented
in~\cite{Barenboim2024}, our results suggest that the anti-trapped
region may have a longer lifetime in our scenario, at least in the
absence of additional perturbations. Outgoing geodesics in the
interior of this anti-trapped region also appear to experience an
exponential slow-down of their affine parameters, in a way akin to the
interior of a mass-inflating classical black
hole~\cite{Brady1995,Barcelo2022}.

The paper is organised as follows. In section~\ref{s2} we introduce
the details of the semiclassical system we will work with, presenting
the equations and initial conditions we will use. We also briefly
discuss the numerical scheme we employ. In section~\ref{s3} we present
the results of our analysis, involving the evaporation of the trapped
region and the appearance of the anti-trapped region beneath it. In
section~\ref{s4} we put our result in the broader context of
inner-horizon-related physics and discuss possible future avenues of
research which would further improve our understanding of the
evolution of BHs in general.

\section{Charged black hole and semiclassical gravity}\label{s2}

Throughout this paper we will work with 3+1 dimensional spherically
symmetric geometries. Additionally, for the purpose of simplifying the
expressions of the semiclassical terms we introduce below, we use
double null coordinates for the radial-temporal sector. The line
element of our geometry is then
\begin{equation}\label{geo}
	ds^2=-C(u,v)dudv+r(u,v)^2(d\theta^2+\sin^2\!\theta\,d\phi^2),
\end{equation}
where $C$ is the \textit{conformal factor} of the radial-temporal
sector, and $r$ is the \textit{areal radius}. For reference, the
non-zero components of the Einstein tensor of this metric in this
coordinate system are given by
\begin{equation}
	\begin{split}
		G_{uu}&=\frac{2C_{,u}r_{,u}}{Cr}-\frac{2r_{,uu}}{r},\\
		G_{vv}&=\frac{2C_{,v}r_{,v}}{Cr}-\frac{2r_{,vv}}{r},
	\end{split}
\end{equation}
and
\begin{equation}
	\begin{split}
		G_{uv}=G_{vu}&=\frac{C}{2r^2}+\frac{2r_{,u}r_{,v}}{r^2}+\frac{2r_{,uv}}{r},\\
		G_{\theta\theta}=\frac{G_{\phi\phi}}{\sin^2\!\theta}&=\frac{2r^2C_{,u}C_{,v}}{C^3}-\frac{2r^2C_{,uv}}{C^2}-\frac{4rr_{,uv}}{C},
	\end{split}
\end{equation}
where a comma denotes partial differentiation.

\subsection{Classical electromagnetic field}

For the (effectively) classical stress-energy content we consider a
spherical electric field $E(u,v)$, sourced by a point charge at the
origin. The corresponding Faraday tensor in null coordinates has the
non-zero components
\begin{equation}
	\begin{split}
		F_{uv}=-F_{vu}=\frac{C}{2}E.
	\end{split}
\end{equation}
The stress-energy tensor (SET) of this field, given by
\begin{equation}
	T_{\mu\nu}=F_{\mu\rho}F_{\nu\sigma}g^{\rho\sigma}-\frac{1}{4}g_{\mu\nu}F_{\rho\sigma}F^{\rho\sigma},
\end{equation}
has the non-zero components
\begin{equation}
	\begin{split}
		T_{uv}=T_{vu}&=\frac{C}{4}E^2,\\
		T_{\theta\theta}=\frac{T_{\phi\phi}}{\sin^2\!\theta}&=\frac{r^2}{2}E^2.
	\end{split}
\end{equation}
Raising one index, the tensor takes on the familiar form from the
Reissner-Nordström solution,
\begin{equation}
	(T_{\mu}^{\phantom{\mu}\nu})=\frac{E^2}{2}\begin{pmatrix}
		-1&0&0&0\\0&-1&0&0\\0&0&1&0\\0&0&0&1
	\end{pmatrix}.
\end{equation}
The conservation and Maxwell equations then give
\begin{equation}
	\begin{split}
		E_{,u}+\frac{2r_{,u}E}{r}=0,\\
		E_{,v}+\frac{2r_{,v}E}{r}=0.\\
	\end{split}
\end{equation}
After a change of function $E(u,v)\to q(u,v)$, defined by
\begin{equation}
	E(u,v)=\frac{q(u,v)}{r(u,v)^2},
\end{equation}
the equations reduce to
\begin{equation}
	q_{,u}=q_{,v}=0,
\end{equation}
which implies that the generic solution is $q=\text{const.}$ This is
simply a consequence of the lack of s-wave dynamics for the vacuum
electromagnetic field. The important point for our purposes is that
this result relies only on the independent conservation of the
classical SET, and will therefore still hold when we include the
semiclassical source into the mix, as long as the quantum field is not
charged. We choose a rescaled charge parameter $Q=q\sqrt{4\pi}$, such
that the electric field component has the form
\begin{equation}\label{emfield}
	E(u,v)=\frac{Q}{\sqrt{4\pi}r(u,v)^2},
\end{equation}
and the classical electrovacuum solution has the standard
Reissner-Nordström form in geometric units ($G=c=1$), which in the
Eddington-Finkelstein null coordinates would be eq.~\eqref{geo} with
\begin{equation*}
	C(u,v)=1-\frac{2M}{r(u,v)}+\frac{Q^2}{r(u,v)^2}.
\end{equation*}
We note that below we will not use the Eddington-Finkelstein $u$, but
rather a different $u$ coordinate adapted to the vacuum state of the
quantum field.

\subsection{Quantum scalar field}

In the present work, we use the Polyakov approximation to the RSET of
a scalar field, which encodes the non-local terms that drive Hawking
evaporation and other horizon-related dynamics, while keeping the full
system of equations in closed form and of second order. The
approximation is obtained for a massless minimally coupled scalar by
considering its propagation on a 1+1 dimensional background
corresponding to the radial-temporal sector of a spherically-symmetric
3+1 geometry. The 1+1 dimensional line element can be written in null
coordinates in the form
\begin{equation}
	ds^2_{(2)}=-C(u,v)dudv.
\end{equation}
Due to the conformal invariance of the Klein-Gordon equation and the
conformal flatness in 1+1 geometries, the mode solutions which one
requires for quantisation can be obtained
analytically~\cite{DaviesFulling},
\begin{equation}\label{planewaves}
	\phi^{(u)}=\frac{1}{\sqrt{4\pi\omega}}e^{-i\omega u},\quad \phi^{(v)}=\frac{1}{\sqrt{4\pi\omega}}e^{-i\omega v},
\end{equation}
where $\omega$ is the frequency. A mode basis of this form can be
written for every pair of null coordinates $\{u,v\}$, generally
leading to non-unitarily-equivalent quantisations. For gravitational
collapse in asymptotically-flat spacetimes, a standard choice is the
``in" quantisation, which recovers a Minkowski quantisation at past
(null) infinity, and is thus considered to have reasonable initial
conditions.

Once a quantisation is defined, a SET operator can be constructed, and
its vacuum expectation value can be computed and covariantly
renormalised. This RSET in two dimensions $\expval{T_{ab}}^{(2)}$ can
then be related to the four-dimensional Polyakov approximation for the
RSET through the expression
\begin{equation}
	\expval{T_{\mu\nu}}=\delta^a_\mu\delta^b_\nu\expval{T_{ab}}^{(2)}.
\end{equation}
The components of this Polyakov RSET, in the coordinate system
$\{u,v\}$ for which the quantum field modes have the plane wave
expressions~\eqref{planewaves}, have the form~\cite{Fabbri2005}
\begin{equation}\label{RSET}
	\begin{split}
	\expval{T_{uu}}&=\frac{l_{\rm p}^2}{96\pi^2r^2}\left(\frac{C_{,uu}}{C}-\frac{3}{2}\frac{C_{,u}^2}{C^2}\right),\\
		\expval{T_{vv}}&=\frac{l_{\rm p}^2}{96\pi^2r^2}\left(\frac{C_{,vv}}{C}-\frac{3}{2}\frac{C_{,v}^2}{C^2}\right),\\
		\expval{T_{uv}}&=\frac{l_{\rm p}^2}{96\pi^2r^2}\left(\frac{C_{,u}C_{,v}}{C^2}-\frac{C_{,uv}}{C}\right),
	\end{split}
\end{equation}
where $l_{\rm p}$ is the Planck length. Depending on the choice of
field modes, or, equivalently, of a vacuum state, the RSET can have
drastically different behaviours. For instance, the Boulware
state~\cite{Boulware1975} on a black hole background is obtained when
$u$ and $v$ in \eqref{planewaves} are the Eddington-Finkelstein
coordinates, and gives a singular behaviour at both past and future
horizons. On the other hand, the Hartle-Hawking
state~\cite{HartleHawking1976} is obtained when $u$ and $v$ are the
Kruskal coordinates regular at the past and future outer horizon.

The ``in" vacuum state can also be obtained with these expressions by
using a $v$ coordinate proportional to the Eddington-Finkelstein $t+r$
(in the asymptotic past), where $t$ is the standard Minkowski time,
and a $u$ coordinate constructed from reflection of ingoing light rays
at the origin, i.e. $u\propto v$ at $r=0$. For simplicity, all results
in subsequent sections will be presented in this coordinate system,
since it will be regular at both the outer and inner apparent horizons
generated in our collapse model.

The Polyakov approximation itself, as related to 3+1 dimensional
computations of field modes and the RSET, boils down to considering
only the $s$-wave dynamics of the field, and disregarding
backscattering from the field potential. Its accuracy is therefore
limited to cases in which backscattering is sufficiently negligible
and non-spherical modes are subdominant in the vacuum energy
contribution. In practice it has been shown to reproduce the Hawking
effect~\cite{DFU1976,Barcelo2021}, and a comparison between
Refs.~\cite{Zilberman2019} and \cite{Barcelo2021} (as well as the
results below) shows that the ingoing flux component on the inner
horizon of a Reissner-Nordström black hole is accurate in a regime
close to extremality. A more quantitative comparison, however, would
require exact results for the four-dimensional RSET in the ``in" state
of a collapse model, which are not yet available in the literature. We
therefore present the below results in the spirit of a qualitative
analysis, which would likely be an accurate description of exact
four-dimensional semiclassical gravity only in some particular regime
of the parameter space.

With the addition of the Polyakov RSET, the semiclassical Einstein
equations describing the Reissner-Nordström system are
\begin{align}
	\frac{2C_{,u}r_{,u}}{Cr}-\frac{2r_{,uu}}{r}&=\frac{L_{\rm p}^2}{r^2}\left(\frac{C_{,uu}}{C}-\frac{3}{2}\frac{C_{,u}^2}{C^2}\right),\label{guu}\\
	\frac{2C_{,v}r_{,v}}{Cr}-\frac{2r_{,vv}}{r}&=\frac{L_{\rm p}^2}{r^2}\left(\frac{C_{,vv}}{C}-\frac{3}{2}\frac{C_{,v}^2}{C^2}\right),\label{gvv}\\
	\frac{C}{2r^2}+\frac{2r_{,u}r_{,v}}{r^2}+\frac{2r_{,uv}}{r}&=\frac{L_{\rm p}^2}{r^2}\left(\frac{C_{,u}C_{,v}}{C^2}-\frac{C_{,uv}}{C}\right)+\frac{Q^2C}{2r^4},\label{guv}\\
	\frac{2r^2C_{,u}C_{,v}}{C^3}-\frac{2r^2C_{,uv}}{C^2}-\frac{4rr_{,uv}}{C}&=\frac{Q^2}{r^2},\label{gthth}
\end{align}
where $Q=\text{const.}$ and $L_{\rm p}^2=l_{\rm p}^2/(12\pi)$, and
where we have used geometric units $G=c=1$. Note that, although it
represents the energy content of the scalar field in its vacuum state,
the RSET is unlike a standard classical SET in that it contains second
derivatives of the metric function $C$, changing the principal part of
the equations. In this sense, these equations are more akin to a
modified gravity theory. We also note that if the calculation of the
RSET were performed in four dimensions, then up to fourth order
derivatives would appear in the equations, and spurious non-physical
solutions would be present~\cite{Horowitz1978}. While this issue could
then potentially be resolved with a perturbative
order-reduction~\cite{Simon1990}, an added advantage of the Polyakov
approximation is that this additional step is not necessary.

For the initial value problem in this coordinate system, it is most
convenient to use equations \eqref{guv} and \eqref{gthth} for the
evolution, and equations \eqref{guu} and \eqref{gvv} as constraints on
the two null initial value surfaces. The former two give the
expressions for the cross-derivatives
\begin{align}
	r_{,uv}&=-\frac{r}{4}\frac{C+4r_{,u}r_{,v}}{r^2-L_{\rm p}^2}+\frac{Q^2C}{4r^3}\frac{r^2+L_{\rm p}^2}{r^2-L_{\rm p}^2},\label{eqr1}\\
	C_{,uv}&=\frac{C_{,u}C_{,v}}{C}+\frac{1}{2}\frac{C^2+4Cr_{,u}r_{,v}}{r^2-L_{\rm p}^2}-\frac{Q^2C^2}{r^2(r^2-L_{\rm p}^2)}.\label{eqc1}
\end{align}
The remaining two, \eqref{guu} and \eqref{gvv}, must be used to choose
adequate initial data.

\subsection{Gravitational collapse}

\begin{figure}[t!]
	\centering
	\includegraphics[scale=0.6]{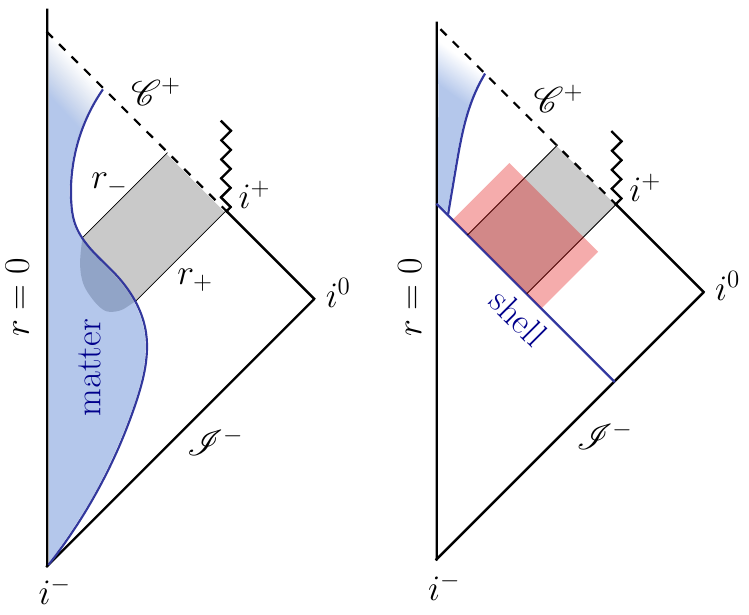}
	\caption{Causal diagram of dynamically formed charged black
          hole. Left: realistic model where a charged fluid (blue
          shaded region) collapses and forms a trapped region (grey
          shaded region). Right: simplified collapse model where the
          matter is initially approximated by a thin shell. The region
          where the evolution is studied numerically (shaded in red on
          the right) is qualitatively similar between the two.}
	\label{f1}
\end{figure}

The causal structure of a dynamically formed Reissner-Nordström black
hole, as studied for example in~\cite{Boulware1973,Krasinski2006}, is
depicted in figure~\ref{f1}. In particular, the left panel of the
figure shows a generic formation scenario with a continuous
distribution of matter. While the asymptotic radial approach of the
surface toward $r_-$ depicted here is from below, we note that it can
also be from above, and this would not change the dynamics of the
inner apparent horizon substantially (this horizon would just tend
toward $r_-$ exponentially quickly).

The right panel of figure~\ref{f1} depicts a simplification of the
collapse scenario which we will use in order to set up initial data
for the ``in" vacuum in the black hole region. This simplification
consists of considering the collapsing matter as compressed into a
light-like thin shell, akin to the Vaidya model, which is often used
in semiclassical
constructions~\cite{DFU1976,Singh2014,Anderson2020}. We recall that
the standard Unruh state here would be singular at the inner apparent
horizon, hence the need to explicitly construct the ``in" state. The
red shaded diamond in the figure is a representation of the domain
where we will evolve the semiclassical equations, which will lead to
modifications in the horizon positions compared to this picture. A
neighbourhood of the origin in the black hole region is depicted as
matter-filled in order to highlight the fact that a timelike
singularity would not typically form from such a collapse, making the
thin shell approximation inadequate close to the would-be
singularity. Our numerical analysis will not extend to the origin
partly because of the added complexity of having to consider a
matter-filled region, and partly because of the need to regularise the
Polyakov approximation if it is used in proximity to the
origin~\cite{Arrechea2021}.

Additionally, we observe that even classically, this causal diagram
would be modified if generic perturbations are
considered. Particularly, in the presence of generic disturbances to
the gravitational and matter fields after black hole formation, the
inner region of this geometry would undergo the mass inflation
instability~\cite{Poisson1989,Ori1991,Dafermos2003,Dafermos2017}. In
the present work we disregard this effect and focus on the purely
RSET-sourced horizon dynamics. Our aim is to provide insight into the
effect that such a semiclassical source can have, particularly
asymptotically, where it is expected to dominate over the classical
instability~\cite{Hollands2019,Zilberman2019}. We will, however,
consider the addition of classical perturbations to this type of
system in a future work.

\subsection{Numerical method and initial data}

The system of equations is solved with a second order Runge-Kutta
method. Reduction to first order is obtained by introducing the
auxiliary variables $s=r_{,u}$, $t=r_{,v}$, $A=C_{,u}$,
$B=C_{,v}$. Due to working with double null coordinates, some of the
variables must be integrated in the $v$ direction and others in the
$u$ direction. From the definition of the reduction variables and
equations \eqref{guu}, \eqref{gvv}, \eqref{eqr1} and \eqref{eqc1}, we
have the system of equations and constraints
\begin{align}
	s_{,v}=t_{,u}&=-\frac{r}{4}\frac{C+4st}{r^2-L_p^2}+\frac{Q^2C}{4r^3}\frac{r^2+L_p^2}{r^2-L_p^2},\label{sev}\\
	A_{,v}=B_{,u}&=\frac{AB}{C}+\frac{1}{2}\frac{C^2+4Cst}{r^2+L_p^2}-\frac{Q^2C^2}{r^2(r^2-L_p^2)},\label{aev}\\
	t_{,v}&=\frac{Bt}{C}-\frac{L_p^2}{2r}\left(\frac{B_{,v}}{C}-\frac{3}{2}\frac{B^2}{C^2}\right),\label{constraintu}\\
	s_{,u}&=\frac{As}{C}-\frac{L_p^2}{2r}\left(\frac{A_{,u}}{C}-\frac{3}{2}\frac{A^2}{C^2}\right),\label{constraintv}\\
	r_{,v}&=t,\label{rev}\\
	C_{,v}&=B,\label{cev}\\
	r_{,u}&=s,\label{sdef}\\
	C_{,u}&=A\label{Adef}.
\end{align}
We choose to evolve $\{r,C,s,A\}$ in $v$, and $\{t,B\}$ in $u$. Given
that these equations include the RSET components written in the null
coordinates corresponding to the mode quantisation
basis~\eqref{planewaves}, they are only accurate in the ``in" null
coordinates $\{u,v\}$, of which $v$ is the Eddington-Finkelstein
ingoing coordinate and $u$ is obtained from $v$ through the reflection
condition ot the origin $u|_{r=0}=v|_{r=0}$.

Without loss of generality, we choose the ingoing shell depicted in
the right panel of fig.~\ref{f1} to be located at $v=0$. In the
absence of other matter sources, the initial conditions at this
surface for the functions evolved in $v$ become
\begin{equation}\label{condv1}
	r(u,0)=-\frac{u}{2},\quad C(u,0)=-2r_{,u}|_{v=0}=1,\quad s(u,0)=r_{,u}|_{v=0}=-\frac{1}{2},\quad A(u,0)=C_{,u}|_{v=0}=0.
\end{equation}
The first two of these conditions are obtained considering that the
spacetime is Minkowski before $v=0$, while the last two are obtained
from constraints which define the order-reduction variables. One can
also easily check that these conditions are compatible with the
additional constraint at this surface given by
eq.~\eqref{constraintv}, since the $uu$ component of the RSET on this
surface is the same as in Minkowski spacetime, which is
zero.\footnote{Note that one of the requirements for constructing a
  viable RSET is that flat spacetime be a solution to the
  semiclassical vacuum equations~\cite{Wald1995}, implying this
  condition goes beyond our particular choice of RSET prescription,
  when the appropriate vacuum state is chosen.}

Choosing the initial conditions for $\{t,B\}$ on the outer $u$ surface
is a bit more subtle. Since Reissner-Nordström is not a solution of
the semiclassical equations, and we do not have an analytical
expression for any particular such solution, we are left with two
options. We either compute the full self-consistent solution all the
way back to past null infinity, or we make a choice for its expression
on a finite $u$ surface which we expect to be similar enough to this
exact solution. Here we opt for the second choice, and we fix $C$ on
this $u=u_0$ surface to be the same as in the classical
Reissner-Nordström geometry, subsequently obtaining $r$ by integrating
the semiclassical constraint~\eqref{constraintu}. In practice this
does lead to a semiclassical solution on this surface, but one in
which the past evolution has a slight deviation from the shell
collapse model described, or, equivalently, one in which the ``in"
state is slightly deformed. However, since this deformation only
corresponds to a deviation of the solution outside the black hole
region, where semiclassical corrections are small, we expect that this
will lead to no qualitative changes. As to why we fix $C$ to the
classical solution rather than $r$, there are two reasons. On the one
hand, the RSET on $u_0$ will have the same values as on the classical
background solution, which guarantees we have a control over its
magnitude. On the other hand, we avoid a division by $L_p^2$ which
would appear in an equation for $B_{,v}$, and could easily amplify the
deviation away from the semiclassically corrected black hole solution
we seek.

We can express the classical geometry on $u_0$ in advanced
Eddington-Finkelstein coordinates as
\begin{equation}
	ds^2=-f(\tilde{r})dv^2+2dvd\tilde{r}+\tilde{r}^2d\Omega^2,\qquad f(r)=1-\frac{2\tilde{M}}{\tilde{r}}+\frac{Q^2}{\tilde{r}^2},
\end{equation}
where the $\tilde{r}$ coordinate here is just an auxiliary variable we
use to construct the classical $C(u_0,v)$, since the actual areal
radius of the semiclassical solution will be calculated later with
\eqref{constraintu}. Switching to ``in" null coordinates involves
solving the null geodesic equation
\begin{equation}\label{ruv}
	\frac{d\tilde{r}}{dv}=\frac{1}{2}f(\tilde{r}),\qquad \tilde{r}(v=0)=-\frac{u}{2},
\end{equation}
from which one obtains $\tilde{r}(u,v)$, and the line element can be
transformed to the form~\eqref{geo} identifying the function
\begin{equation}
	C(u,v)=-2\frac{\partial\tilde{r}}{\partial u}.
\end{equation}
For the case of Reissner-Nordström, eq.~\eqref{ruv} can be integrated
to give an implicit solution for $\tilde{r}(u,v)$, from which one can
obtain $\tilde{r}(u_0,v)$ with a numerical root finding procedure (in
particular we found that a bisection method works best, due to the
divergent nature of the logarithms in the implicit relation at the
horizon). The conformal factor then is
$C(u_0,v)=f(\tilde{r}(u_0,v))/f(\tilde{r}=-u_0/2)$, and its $v$
derivatives ($B$ and $B_{,v}$) are obtained with the relation
$\partial\tilde{r}/\partial v=f(\tilde{r}(u_0,v))/2$. With all these
expressions, eq.~\eqref{constraintu} can then be integrated to give
$r(u_0,v)$.

With these expressions for the initial data, all constraints are
satisfied, and all free functions are fixed to resemble a classical
Reissner-Nordström black hole. The only free parameters to choose are
the Planck scale $L_p$ and the mass and charge of the black hole or,
equivalently, the radial positions of the outer and inner horizons
$r_\pm$ in the classical configuration. In the numerical
implementation we will use units $r_+=1$, and we will fix $r_-$ to
some value in between $L_p$ and 1. The lower and upper boundaries of
the domain in $u$ will be chosen through the relation $r(u,0)=-u/2$
such that the domain encloses $r_\pm$, but does not go all the way
down to $r=L_p$, where some of the equations become singular. The
domain in $v$ will simply start at $v=0$ and extend up to the maximum
value for which the numerical resolution can produce precise
results. We note that the horizon positions will evolve, and even on
the $v=0$ surface they will not coincide with $r_\pm$, since they will
be determined by the integration in $u$ of $t$ given by
eq.~\eqref{sev}, which already contains semiclassical corrections.

The numerical integration will be done with the second order Heun's
method, which avoids the need to evaluate the functions at midpoints
between adjacent points of the grid, which in turn simplifies the
simultaneous integration in the two directions $u$ and $v$. The
details of the code are briefly summarised in Appendix \ref{AppA}.

\section{Semiclassical evolution}\label{s3}

\subsection{Black hole evaporation}

Our model has two free parameters, which we will fix through the ratio
between the rescaled Planck length and the classical outer horizon
position $L_p/r_+$, and the ratio between the classical inner and
outer horizon positions $r_-/r_+$. In the below results we use $r_+=1$
to set the scale. The charge which appears in the evolution equations
is given by $Q = \sqrt{r_- r_+}$. As mentioned above, the
semiclassically corrected positions of the inner and outer apparent
horizons will not coincide with the classical ones even on the initial
slice $v=0$, since the integration of the
constraint~\eqref{constraintv} will shift them. We therefore denote
the positions of these true apparent horizons of the geometry by
$\bar{r}_\pm$.

\begin{figure}[t!]
	\centering
	\includegraphics[width =\textwidth]{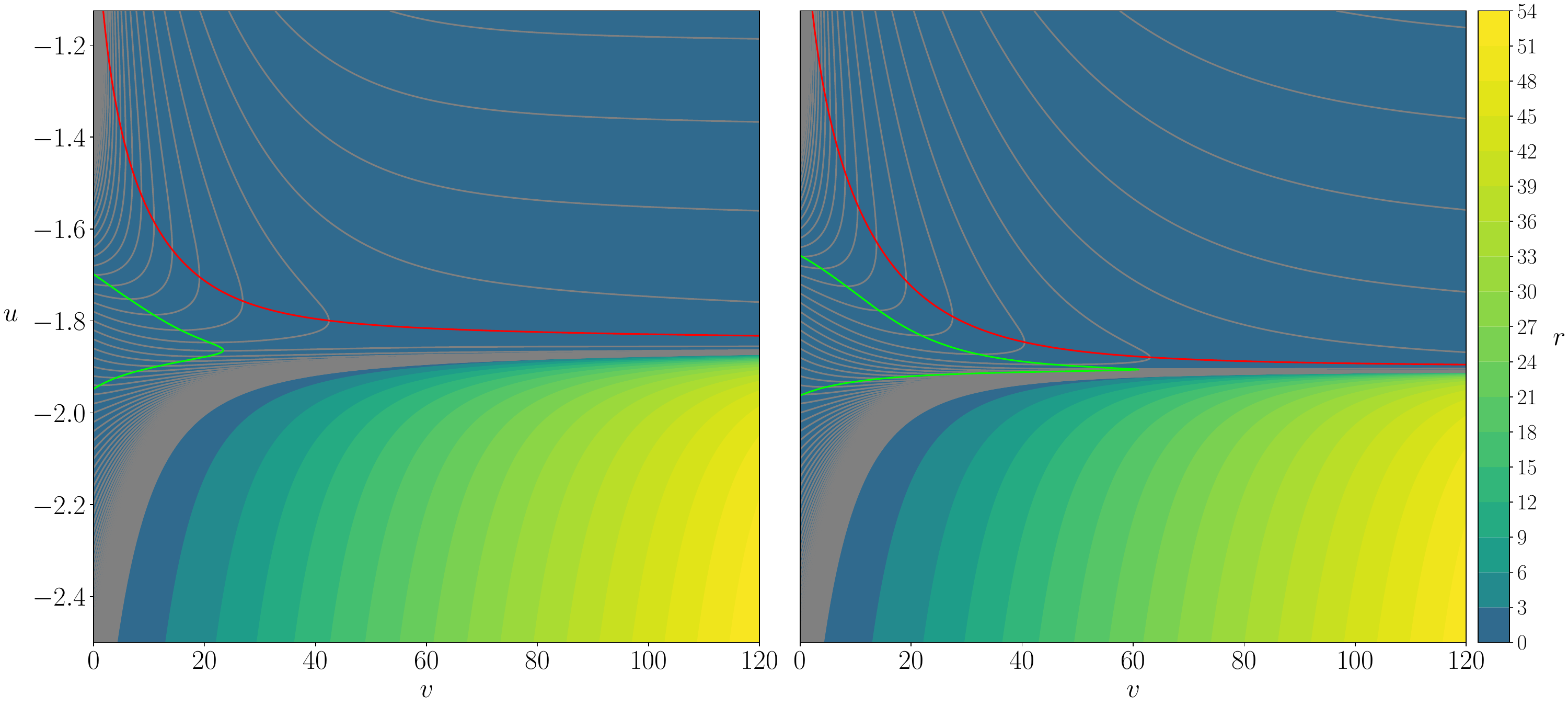}
	\caption{Coloured contour plots for the metric variable
          $r(u,v)$ for two different Plank lengths $L_p$. Superposed
          is a green curve which depicts the boundary of the trapped
          region, that is, the inner and outer apparent horizons,
          corresponding to $r_{,v} = 0$. Also superposed is a red
          curve depicting the boundary of the anti-trapped region,
          corresponding to $r_{,u} = 0$. The grey lines are contours
          of constant $r$ from $r = 1.5$ to $r = 0.5$ (upper left
          corner) with a $0.1$ spacing. Left: Evolution performed with
          parameters $L_p = 0.29, \, r_- = 0.75$ (in units $r_+ =
          1$). The resulting evaporation time for these parameters is
          $v_E \sim 23.4$. Right: Evolution performed with parameters
          $L_p = 0.26, \, r_- = 0.75$. The evaporation time for these
          parameters is $v_E \sim 61.1$.}
	\label{r_contour}
\end{figure}

The causal properties of the numerically evolved geometry can be read
from a contour plot of the areal radius
$r(u,v)$. Figure~\ref{r_contour} depicts such contour plots for two
different values of $L_p$ for a fixed $r_-$. Marginally trapped
surfaces, where the slope of the $r=\text{const.}$ contours becomes
zero or infinity, are marked in colours. In particular, the boundaries
of the trapped region are marked in green. In both cases, the
semiclassical dynamics makes the outer horizon move inwards in what
can be identified as Hawking evaporation. This movement is accompanied
by an outward displacement of the inner horizon, resulting in the
total evaporation of the trapped region in a finite advanced time $v$.

One important aspect of this result is the absence of an extremal
remnant. After the coalescence of the inner and outer apparent
horizons there are no remaining (marginally-)trapped surfaces for the
rest of the evolution. This can be interpreted as a consequence of the
fact that although evaporation-driving fluxes in the RSET are zero
when evaluated on static extremal
backgrounds~\cite{Anderson1995,Balbinot2004}, a dynamical geometry
going through a single time slice of extremality will generally have a
different behaviour compared to its static counterpart in terms of its
derivatives. Thus it is unsurprising that the evolution pushes through
extremality and onto the complete disappearance of the trapped
region. It also suggests that semiclassical stability analyses of
extremal BHs are incomplete if the effect of background dynamics on
the RSET is not studied as well.

Another important part of the result in figure~\ref{r_contour} is the
appearance of an anti-trapped region below the trapped one. Its
boundary, marked with a red contour, has a radius which grows in
$v$. This growth is initially rapid, but then tends to slow down when
it approaches some finite radius close to the size of the last trapped
surface. The lifetime of this incipient white hole appears to be quite
long, even compared to the evaporation timescale of the trapped
region. However, as we will discuss below, its stability under
additional matter perturbations may be unlikely.

While the overall behaviour for the dynamical evolution for both cases
in figure~\ref{r_contour} is qualitatively the same, i.e. a
black-to-white hole transition, both the time for the evaporation of
the trapped region and the growth rate of the anti-trapped region
depend on the Planck length $L_p$. Let us briefly discuss the former.

\begin{figure}[t!]
	\centering
	\includegraphics[scale=0.3]{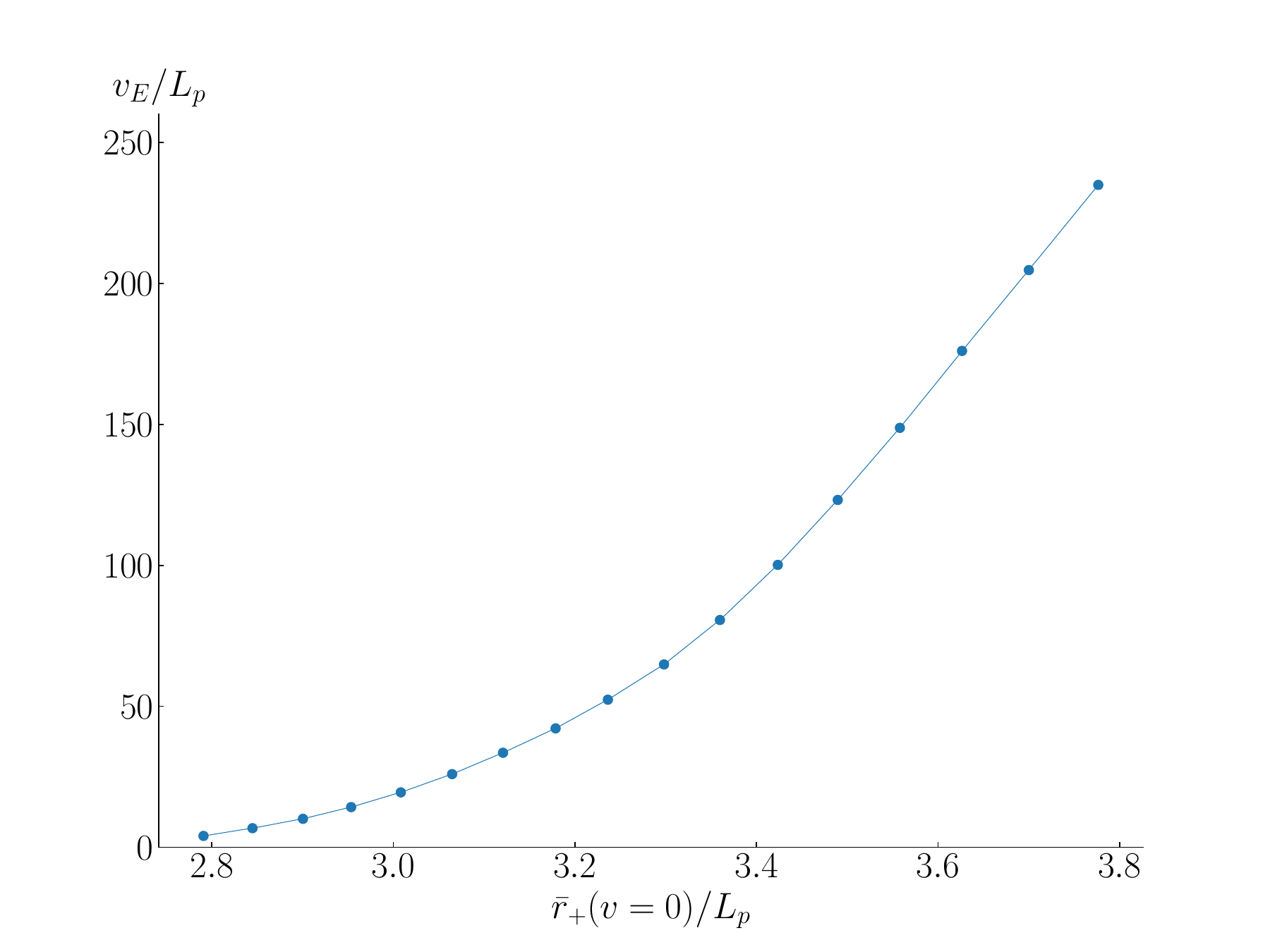}
	\caption{Time of evaporation of the trapped region as a function of the initial outer
          horizon radius $\bar{r}_+$ for different $L_p$ values, from
          $L_p = 0.26$ to $L_p = 0.34$. Both axes are in units of
          $L_p$. Polynomial fits suggest a behaviour which is between quadratic and cubic in $\bar{r}_+$ (the plotted line is simply a first order interpolation between the data points). }
	\label{EvapTimes}
\end{figure}

The trapped region of a Schwarzchild black hole evaporates according
to Hawking's cubic law \cite{Hawking1974}. For a Reissner-Nordström
black hole, even an evaporation driven by the same type of neutral
field is quite different: the (outward) motion of the inner apparent
horizon also must be taken into account. The evaporation therefore
concludes when the two horizons coalesce, rather than when a horizon
reaches a singularity. Figure~\ref{EvapTimes} shows different
evaporation times $v_{E}$ for the trapped region as a function of the
initial semiclassically-corrected outer horizon position
$\bar{r}_+(v=0)$, both divided by $L_p$. We have been able to go up to
a BH size of about 4 times the Planck scale, beyond which the
evaporation times become too long for our numerical method to
convergently resolve that part of the evolution at reasonable
computational cost. How this curve would continue in the regime of
larger scale separations is unclear, particularly due to the fact that
both cubic and quadratic polynomial fits are equally poor for the data
(highly accurate) points available. Analysing the tendency in the
astrophysical regime would therefore require an improved numerical
scheme, which is currently under development. Qualitatively, however,
we can say that the tendency appears to be somewhere between between
quadratic and cubic.

Another interesting aspect of the result visible in
figure~\ref{r_contour} is the sparseness of the $r=\text{const.}$
contours inside the anti-trapped region. This is not an artefact of
the chosen spacing between the lines, but in fact reflects what
appears to be a very slow rate of change, as measured by the ``in"
null coordinates, in the radial position of light rays (and geodesics
in general) in this region. Of particular interest are the outgoing
($u=\text{const.}$) null rays, since this rate of change appears to be
particularly slow in their direction.

\subsection{Causal structure}

\begin{figure}[t!]
	\begin{minipage}{0.5\textwidth}
		\includegraphics[width=\textwidth]{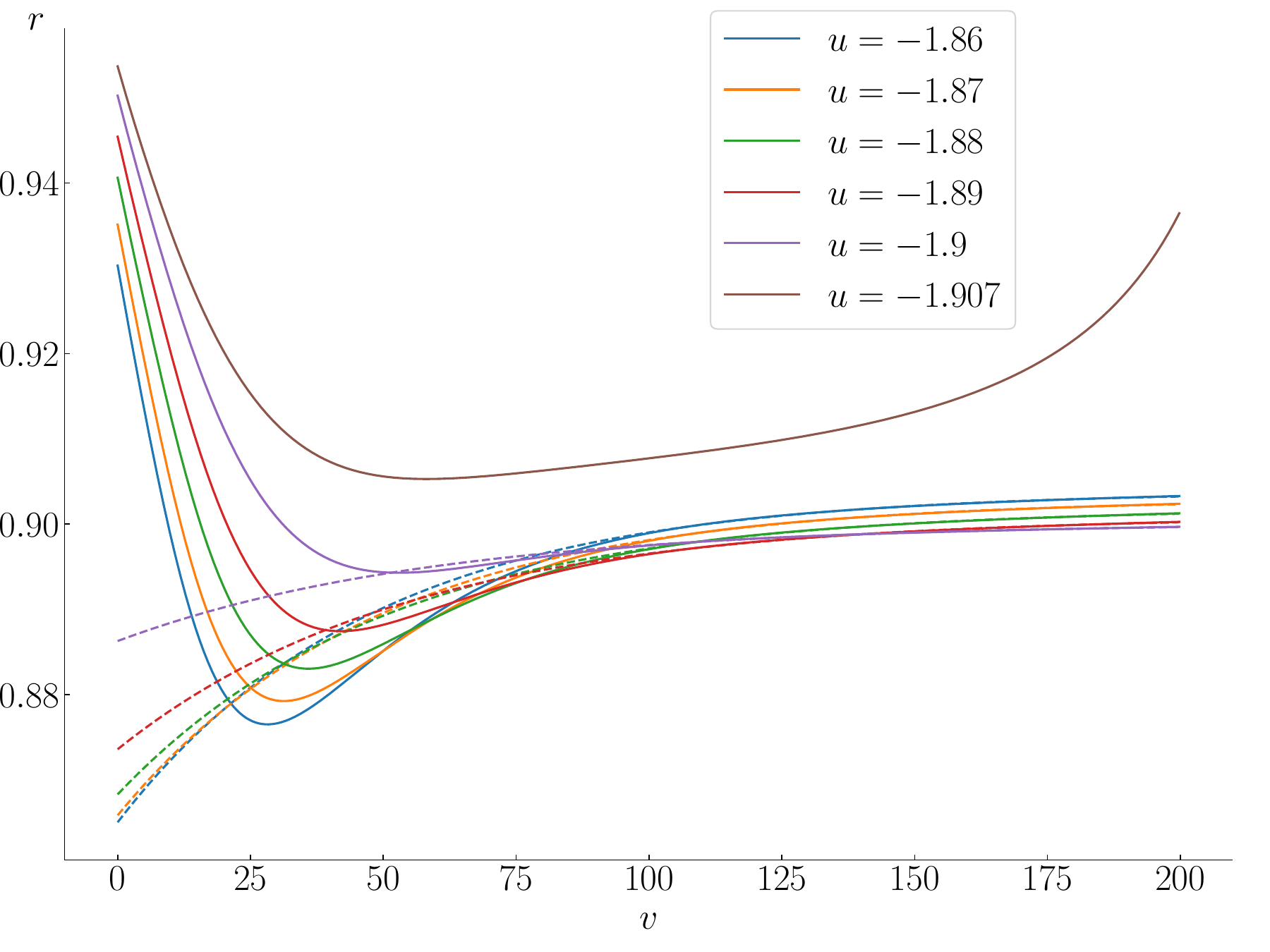}
	\end{minipage}\hfill
	\begin{minipage}{0.5\textwidth}
		\includegraphics[width=\textwidth]{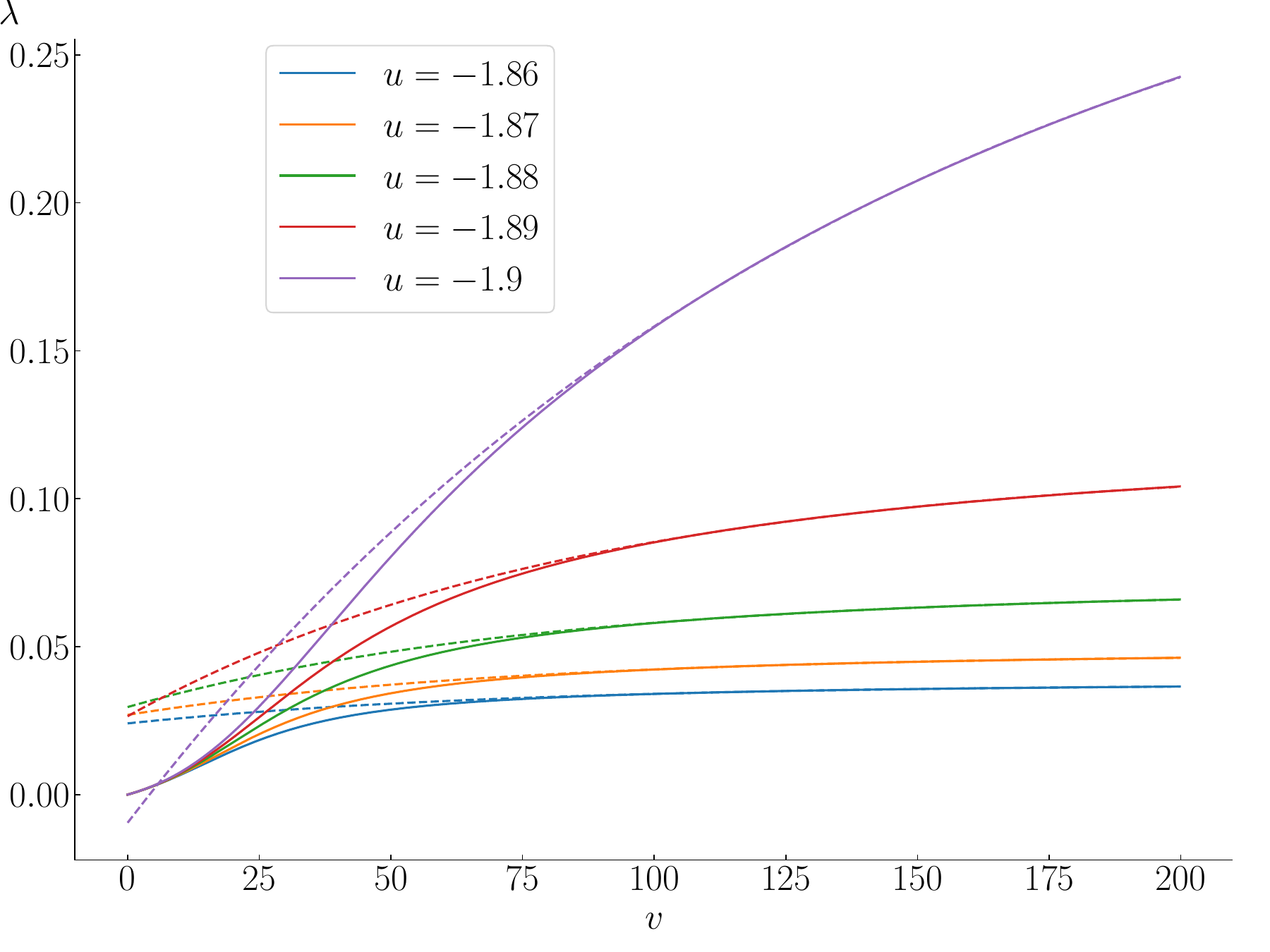}
	\end{minipage}
	\caption{Curves for constant $u$ slices of the affine
          parameter $\lambda(v)$ and $r(v,u_i)$, with
          $u_i = \{-1.86,-1.87,-1.88,-1.89,-1.9,-1.907\}$. Curves that
          remain inside the anti-trapped region are accompanied by an
          inverse exponential fit, depicted by a dashed line of the
          same colour. Left: Curves for the affine parameter $\lambda$
          as a function of $v$. Right: Curves for $r(v,u_i)$ as a
          function of $v$ (note the absence of the $u=-1.907$ curve, for which $\lambda$ would not tend have the same tendency to a constant). The exponential fits show a behaviour reminiscent of the interior of a BH undergoing classical mass inflation.}
	\label{LightRays}
\end{figure}

The left panel of figure~\ref{LightRays} shows slices of the
$L_p = 0.26$ contour plot from figure~\ref{r_contour} for particular
values $u=\text{const.}$ which initially lie inside the trapped
region, then manage to escape it due to its evaporation, to then enter
the anti-trapped region (except the one at $u = -1.907$, which escapes
to infinity). In the figure this is seen as an initial decrease in
their radial position, corresponding to when these light rays are
inside the trapped region, followed by an increase in radius which,
perhaps unexpectedly, becomes slower when they enter the anti-trapped
region. This ``slow down" is in comparison to light rays which reside
outside the anti-trapped region, such as the ray at $u = -1.907$
depicted in the figure, which escapes the trapped region and
subsequently moves out away from the white hole, presumably in an
ordinary expansion toward future null infinity.

\begin{figure}[t!]
	\centering
	\includegraphics[width=0.5\textwidth]{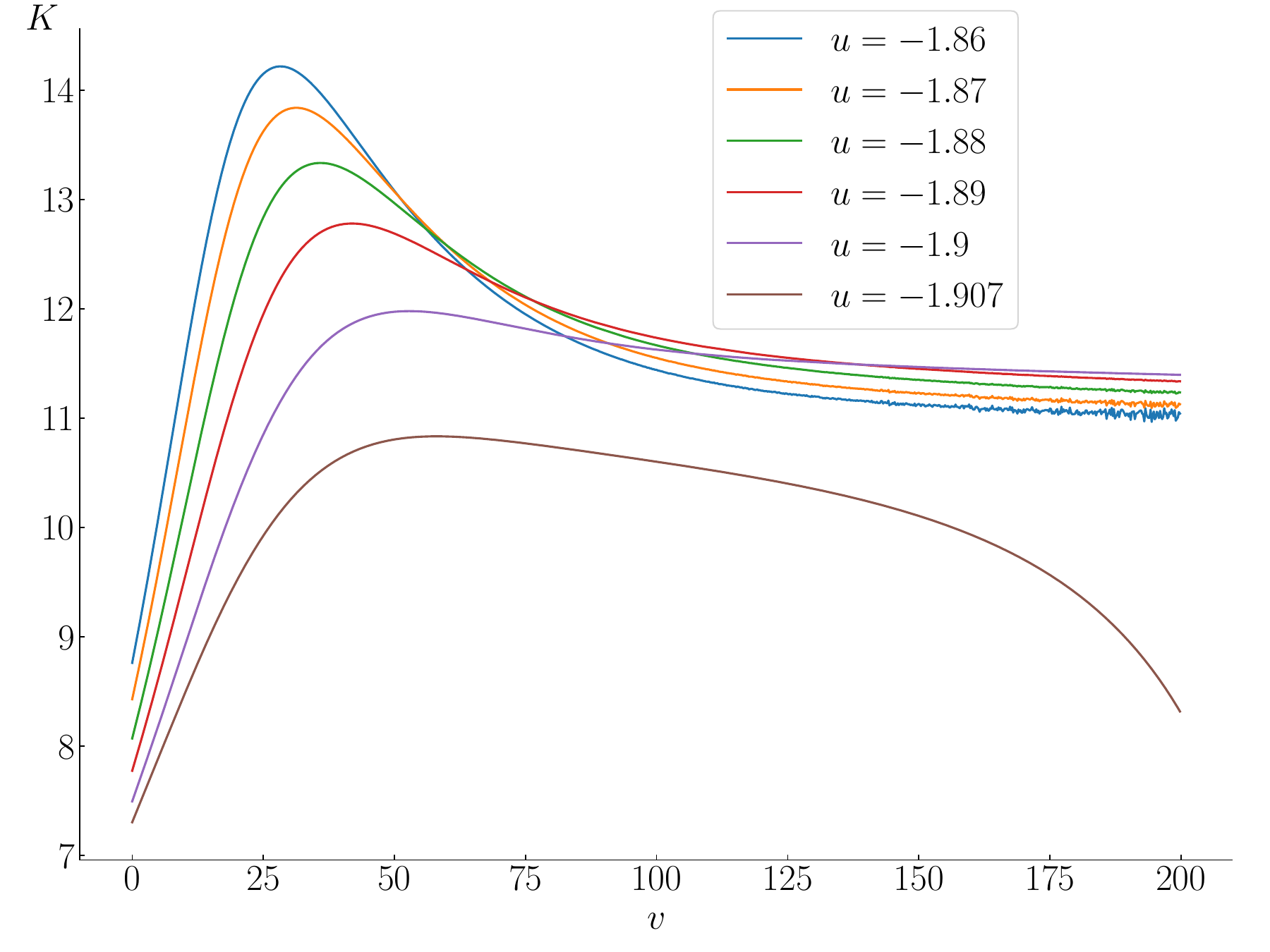}
	\caption{Curves for constant $u$ slices of the Kretschman
          scalar $K$, with
          $u_i = \{-1.86,-1.87,-1.88,-1.89,-1.9,-1.907\}$. The non-growing nature of the curvature is in stark contrast to the classical mass inflation case.}
	\label{AffineParam_KS}
\end{figure}

As this radial slow-down is reminiscent of the trajectories of light
rays inside the trapped region during mass
inflation~\cite{Brady1995,Barcelo2022}, it is curious to check if the
behaviour of the affine parameter $\lambda$ along these geodesics is
in any way similar as well. The geodesic equation which relates the
affine parameter and the advanced time $v$ on $u=\text{const.}$ slices
is
\begin{equation}
	\ddot{v}(\lambda) + \frac{C_{,v}}{C} \dot{v}^2(\lambda) = 0.
\end{equation}
By inverting the above expression, we get a differential equation in
$\lambda(v)$ which we integrate numerically and get the results
represented in the right panel of figure~\ref{LightRays}. Both the
affine parameter $\lambda$ and the radial position $r$ along these
geodesics can be fitted very accurately as a decaying exponential of
the form $a \exp(-b v) + c$ at late times, with $a$, $b$ and $c$
constants. For the affine parameter, if this tendency were to
continue, it would be indicative of the presence of a Cauchy horizon,
since the geodesics would be incomplete at $v\to\infty$. Additionally,
and unlike classical mass inflation, the curvature seems unlikely to
diverge in this limit, as can be seen from extrapolating the behaviour
observed in figure~\ref{AffineParam_KS} in which, after the
evaporation at $v_E \sim 23.4$, the Kretschmann scalar $K$ appears to
steadily decrease along outgoing null rays inside the anti-trapped
region.

\subsection{Curvature analysis}

\begin{figure}[t!]
	\centering
	\includegraphics[scale=0.35]{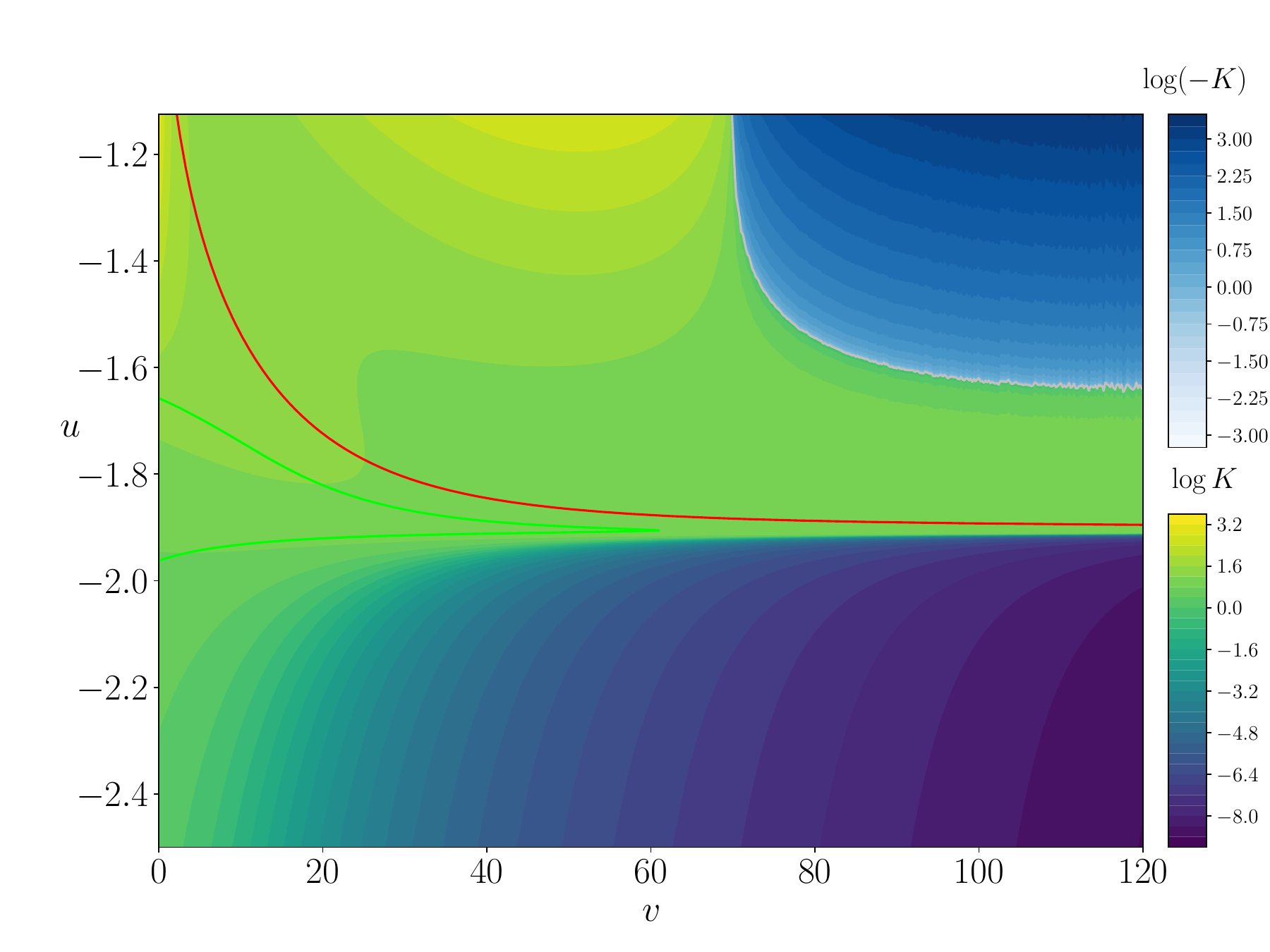}
	\caption{Coloured log scaled contour plot for the Kretschman
          scalar $K$ with $L_p = 0.26$. The contour plot contains two
          different coloured regions adapted to positive and negative
          values. Superposed are a green and a red curve, depicting
          the boundary of the trapped and anti-trapped regions
          respectively.  A grey line is separating the negative and
          positive values regions.}
	\label{KS}
\end{figure}

The Kretschmann scalar for a metric of the form \eqref{geo} reads
\begin{equation}
	\begin{split}
		K = &\frac{4}{r^4} + \frac{16 \, C_{,v}^2 \, C_{,u}^2}{C^6} + \frac{32 \, r_{,v} \, r_{,u}}{C \, r^4}
		+ \frac{64 \, C_{,v} \, r_{,v} \, C_{,u} \, r_{,u}}{C^4 \, r^2} - \frac{64 \, r_{,vv} \, C_{,u} \, r_{,u}}{C^3 \, r^2} + \frac{64 \, r_{,v}^2 \, r_{,u}^2}{C^2 r^4} \\
		& - \frac{32 \, C_{,v} \, C_{,u} \, C_{,uv}}{C^5}
		+ \frac{16 \, C_{,uv}^2}{C^4} + \frac{64 \, r_{,uv}^2}{C^2 r^2} - \frac{64 \, C_{,v} \, r_{,v} \, r_{,uu}}{C^3 r^2}
		+ \frac{64 \, r_{,vv} \, r_{,uu}}{C^2 r^2}.
	\end{split}
\end{equation}
As an example of what the curvature scale is for the geometries
resulting from our numerical evolutions, figure~\ref{KS} shows the
(log-scaled) values of $K$ for our reference example of $r_i=0.75$ and
$L_p = 0.26$. For small values of $v$, $K$ has larger values close to
the origin, $r \to 0$, as one might expect from a Reissner-Nordström
BH. The upper $u$ boundary of our domain, however, has a growing
radius, being initially outside (below) the trapped region, and even
enters the anti-trapped region subsequently. The curvature there
initially decreases, followed by a slight increase, which in turn is
followed by a change to negative values at around $v \sim 70$. At
large $v$ the contour lines of $K$ inside the entire anti-trapped
region seem to tend to become horizontal, much like the
$r=\text{const.}$ lines do. This is also the point observed in
figure~\ref{AffineParam_KS}, namely that outgoing null geodesics
inside this anti-trapped region tend to be locked in an approach
toward a finite curvature and radius, in addition to an infinitely
slowed-down affine parameter.

This last observation is in stark contrast to what the result of
classical mass inflation would be, where curvature would grow along
outgoing null geodesics as they approach the Cauchy horizon, leading
to the creation of a (weak) curvature singularity
there~\cite{Ori1991}. The possible lack of such a growth in this case,
at least for part of the anti-trapped region, would seem to suggest
that this configuration is more similar to a pre-mass-inflation
geometry with a Cauchy horizon. In other words, if we were to perturb
it with an additional (classical) matter source, we might observe a
rapid destabilisation of this seemingly steady asymptotically
approached configuration. This, however, will be the subject of a
future analysis.

\subsection{Polyakov RSET components}

\begin{figure}[th!]
	\centering
	\begin{subfigure}{\textwidth}
		\vspace{-1.3cm}
		\centering
		\includegraphics[width=0.65\textwidth]{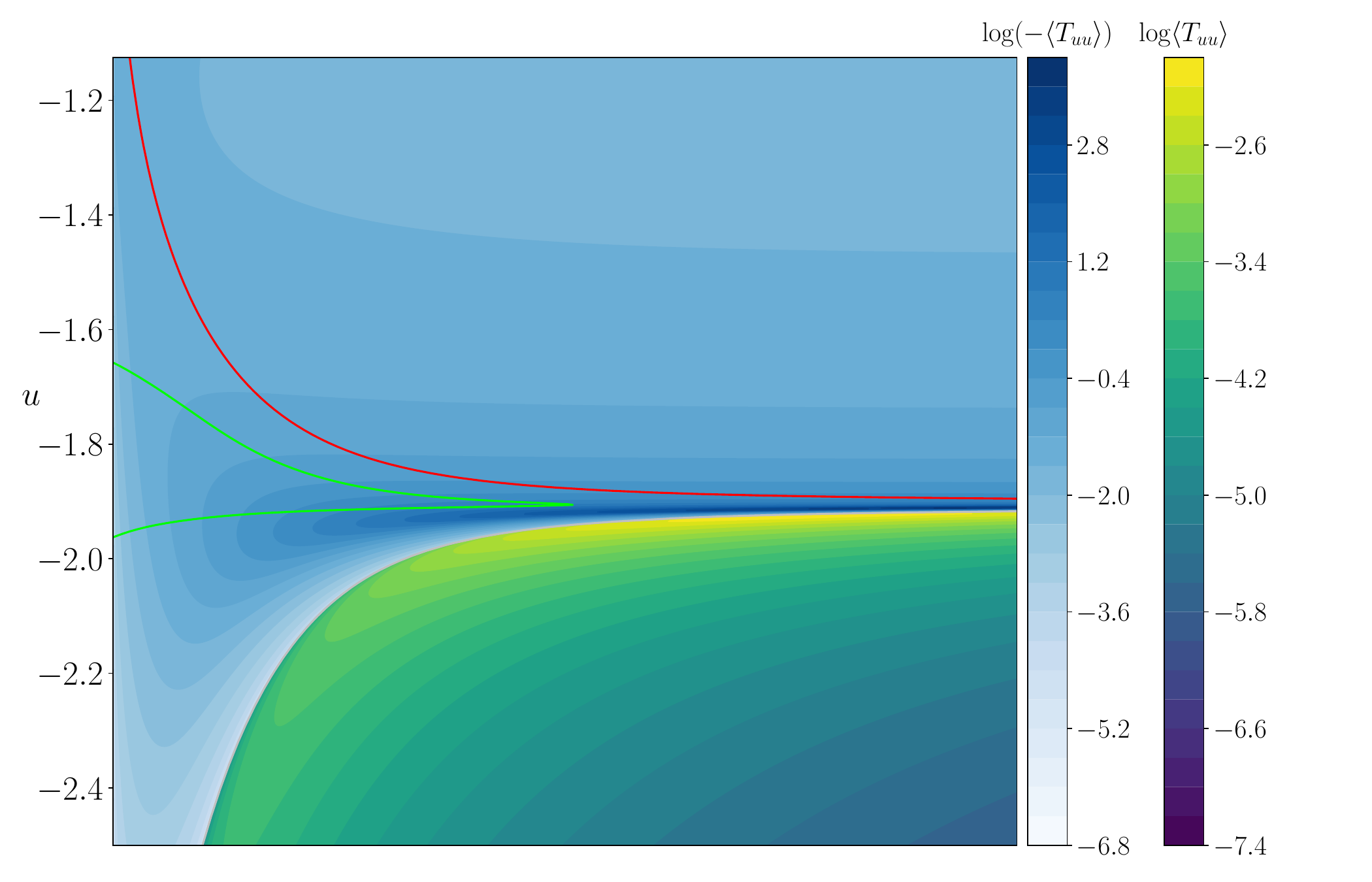}
	\end{subfigure}
	\begin{subfigure}{\textwidth}
		\centering
		\includegraphics[width=0.65\textwidth]{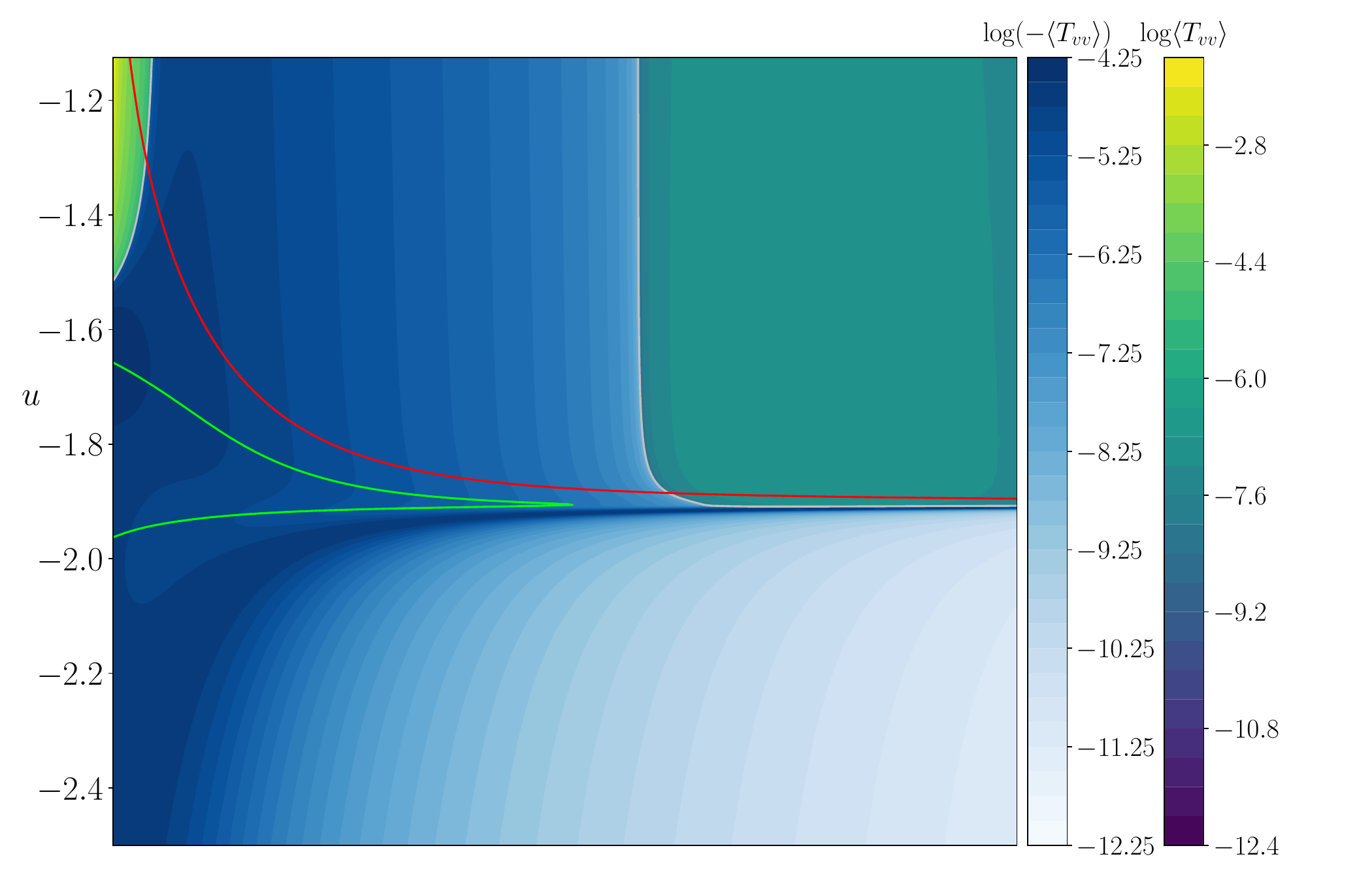}
	\end{subfigure}
	\begin{subfigure}{\textwidth}
		\centering
		\includegraphics[width=0.65\textwidth]{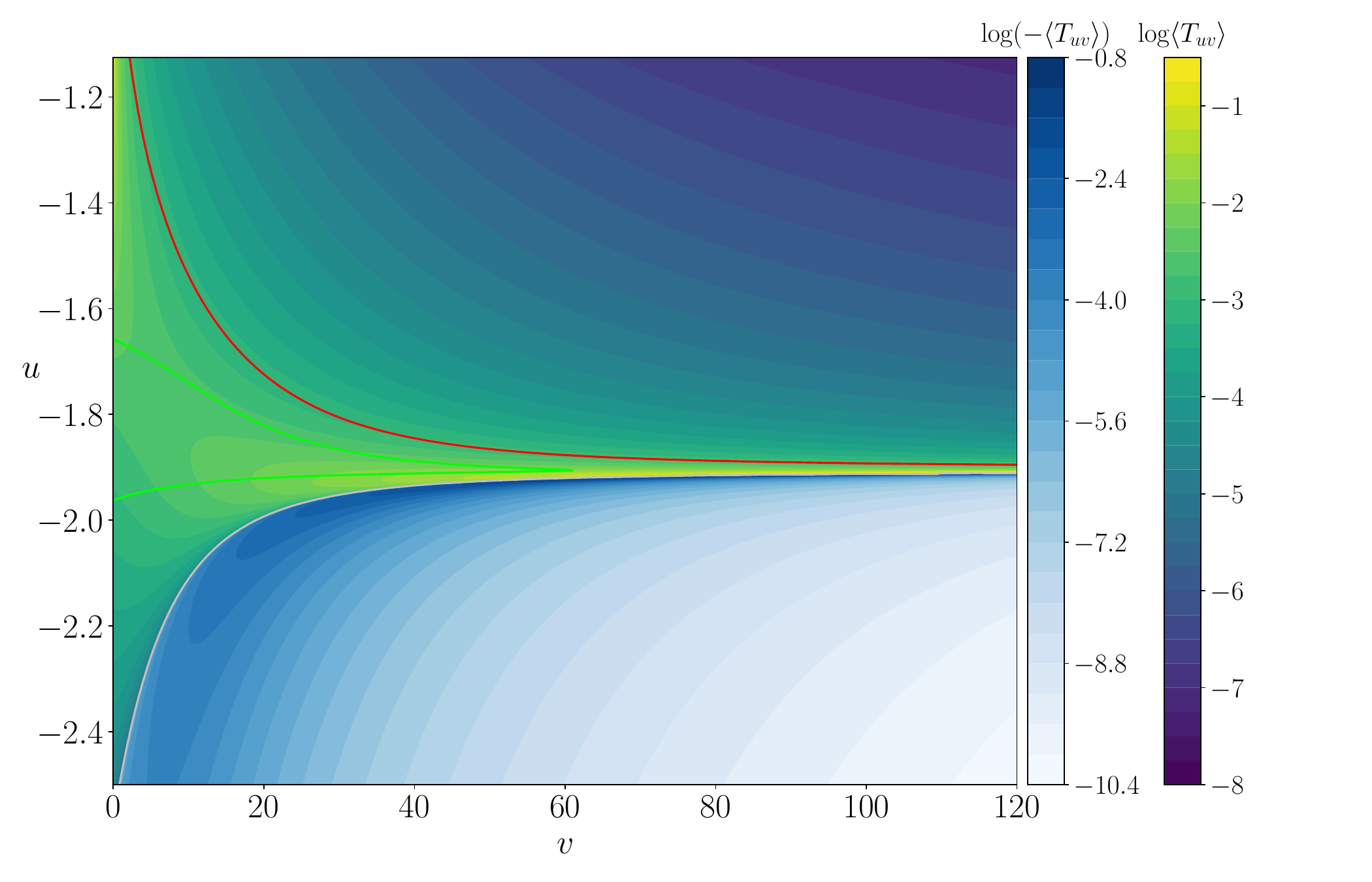}
	\end{subfigure}
	\caption{Coloured log scaled contour plot for the Polyakov
          RSET components with $L_p = 0.26$. The contour plots contain
          two different coloured regions adapted to positive and
          negative values. Superposed are a green and a red curve,
          depicting the boundary of the trapped and anti-trapped
          regions respectively.  A grey line is separating the
          negative and positive values regions.}
	\label{fig:RSET_Components}
\end{figure}

The contour plots of the Polyakov RSET components evaluated in the
numerical integration region are represented in
figure~\ref{fig:RSET_Components} for our reference example of
$L_p=0.26$ and $r_-=0.75$. The upper panel is a plot of the outgoing
flux component $\expval{T_{uu}}$. At initial times $v$, this quantity
is negative for all values of $u$. Then, as more of the region
external to the black hole enters into the integration domain, and as
the ``in" state relaxes to the Unruh state~\cite{DFU1976}, we observe
the positive outgoing flux which goes along with Hawking
evaporation. On the other hand, the flux is negative everywhere else,
and a particularly concentrated burst of it can be found around the
outer horizon of the trapped region toward the end of its
evaporation. The magnitude of this concentrated flux then continues
growing, being the maximum of the contour plot, and its radial
position slowly increases, reaching $r\sim 2$ toward the end of the
evolution.

The middle panel of figure~\ref{fig:RSET_Components} shows the ingoing
flux component $\expval{T_{vv}}$. On the initial surface $v=0$, this
quantity is positive for high values of $u$, below the trapped region,
and negative for $u \lesssim -1.5$, with its minimum value around the
initial inner apparent horizon $\bar{r}_{-}|_{v=0}$ of the trapped
region. This flux has a direct relation to the dynamics of both the
inner and outer apparent horizons, as it can be related to their
initial rates of displacement~\cite{Barcelo2021}. Particularly, a
negative flux at the outer horizon implies this horizon will radially
contract inwards, as occurs during Hawking
evaporation~\cite{DFU1976,ChristensenFulling}, and a negative flux at
the inner horizon implies that this horizon will radially expand
outwards~\cite{Barcelo2021}. Both of these behaviours are indeed
observed in the subsequent dynamics. The flux then continues to be
negative throughout the evaporation process, only switching to a
positive sign around and inside the anti-trapped region after the full
evaporation of the trapped region.

\begin{figure}[th!]
	\centering
	\includegraphics[scale=0.35]{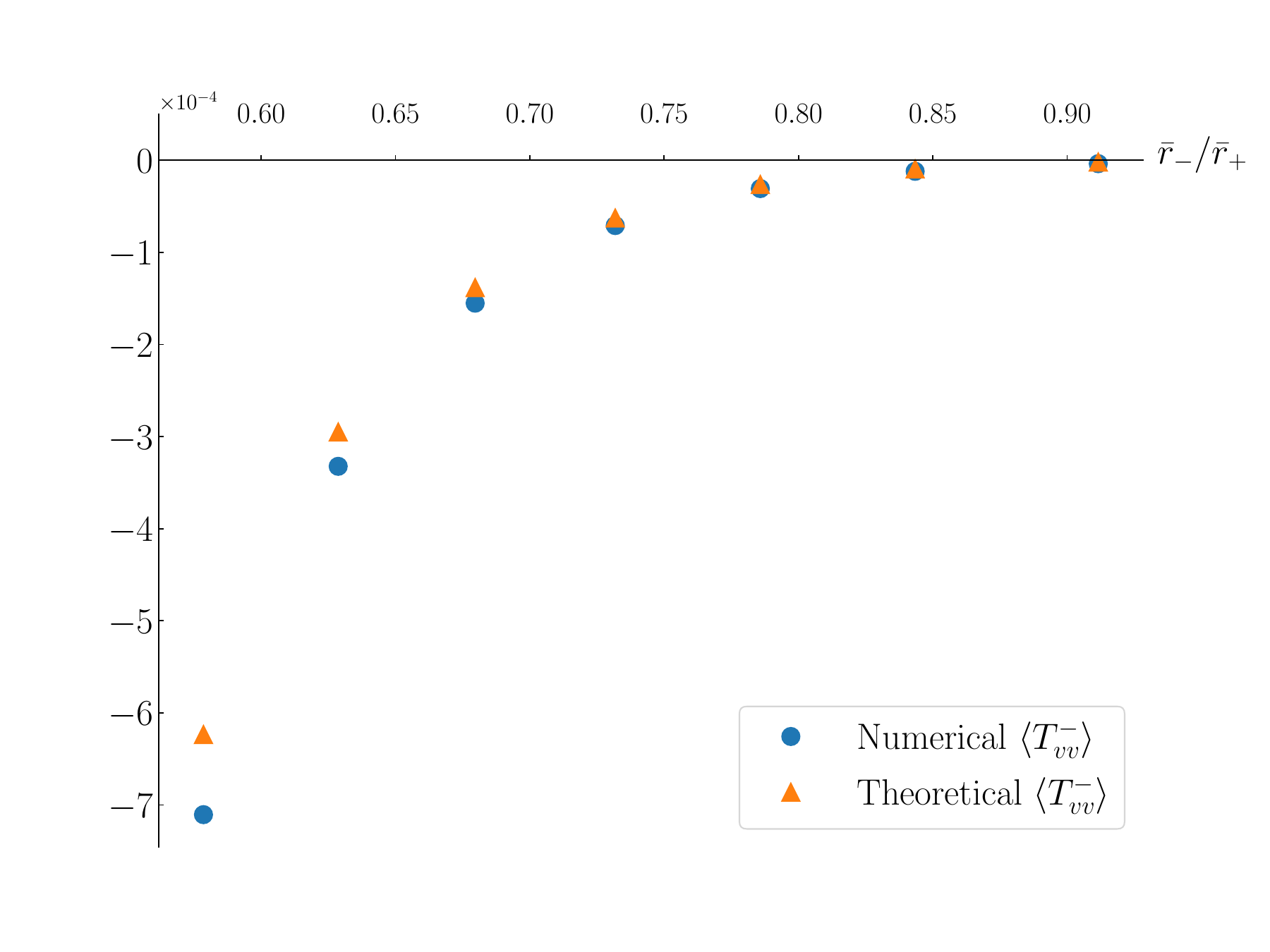}
	\caption{$\expval{T^-_{vv}}$ as a function of the quantum
          corrected $\bar{r}_-/\bar{r}_+$.  Blue data is the RSET
          value at the initial quantum corrected inner horizon at
          $v=0$. Orange data is the theoretical value for the Polyakov
          approximation computed with the quantum corrected initial
          inner and outer horizons.}
	\label{Tvv_CauchyHorizon}
\end{figure}

It is worth making some additional remarks regarding the ingoing flux
at the initial inner horizon position,
$\expval{T_{vv}^-}\equiv\expval{T_{vv}}|_{\{v=0,r=\bar{r}_-\}}$. The
first observation we can make is the fact that $\expval{T_{vv}^-}$
differs slightly from the Polyakov RSET for a static BH with an inner
horizon after the same formation mechanism of a collapsing shell, the
expression for which can be found in equation (33a) of
ref.~\cite{Barcelo2021}. Figure~\ref{Tvv_CauchyHorizon} showcases this
difference by comparing the numerically computed $\expval{T_{vv}^-}$
from our evolved spacetimes with this theoretical computation with no
backreaction, as a function of $\bar{r}_-/ \bar{r}_+$ for a fixed
value of $L_p=0.163$. The cause of the difference is the fact that
$\expval{T_{vv}^-}$ depends not only on the geometry on a constant
time slice, but also on its derivatives. Indeed, for smaller values of
$\bar{r}_-/ \bar{r}_+$ the difference between the numerical and
theoretical $\expval{T_{vv}^-}$, as the flux itself being larger leads
to larger derivatives of the self-consistent background. On the other
hand, closer to an initially extremal configuration, the difference
diminishes. This difference in the derivatives is also what makes the
evaporation push through extremality and onto the full disappearance
of the trapped region, as commented above.

The last panel of figure~\ref{fig:RSET_Components} shows the
$\expval{T_{uv}}$ component of the Polyakov RSET. This component is
related to the Ricci scalar $\mathcal{R}^{(2D)}$ of the induced
two-dimensional metric of the radial-temporal sector, where the field
is quantised~\cite{DaviesFulling}. Particularly, it has the expression
\begin{equation*}
\expval{T_{uv}}=-\frac{l_p^2}{384\pi^2r^2}C\mathcal{R}^{(2D)}.
\end{equation*}
While the Ricci scalar of the four-dimensional Reissner-Nordström BH
is zero (since the SET of the electromagnetic field is traceless),
this scalar on the induced two-dimensional metric has the form
\begin{equation*}
\mathcal{R}^{(2D)}=\frac{4M}{r^3}-\frac{6Q^2}{r^4}.
\end{equation*}
On this background, $\expval{T_{uv}}$ would therefore be positive near
the origin and negative farther away, independently of the vacuum
state in which the RSET is evaluated. The plot of this component in
figure~\ref{fig:RSET_Components} shows that this relation to $r$ is
not significantly changed for the full back-reacted geometry.

\section{Discussion}\label{s4}

Since Hawking's original discovery of BH
evaporation~\cite{Hawking1975,Hawking1974}, the end state of this
process has been an open issue, presumably not answerable within the
range of validity of this semiclassical regime. However, that is not
to say that this process exhausts the applicability of the
semiclassical theory for the analysis of BH dynamics. The fact that
the interior of realistic BHs has a structure which is quite different
from that of the simple Schwarzschild solution means that the
semiclassical regime may be applicable for more than just the slow
outer horizon evaporation.

Generic black holes possess an inner apparent horizon, a lower
boundary to their trapped regions. Classically, this horizon is
destabilised by the mass inflation
process~\cite{Poisson1989,Ori1991,Dafermos2017}, leading to the
formation of a weak singularity in place of the corresponding Cauchy
horizon of unperturbed solutions. Semiclassically, the RSET appears to
have an even stronger energetic contribution than the classical
perturbations which lead to mass inflation~\cite{Birrell1978,
  BalbinotPoisson93, Zilberman2019, Zilberman2022, Hollands2019,
  Hollands2020, McMaken2024}, at least when approaching the Cauchy
horizon.

In this work we have analysed the backreaction form such a
semiclassical source: the Polyakov approximation to the RSET of a
massless scalar field in the ``in" vacuum state, on a dynamically
formed Reissner-Nordström black hole. The dynamics of such a system
had only been previously studied perturbatively around the moment of
horizon formation~\cite{Barcelo2021}. The full non-linear evolution
obtained here leads to two key phenomena, both explicitly
semiclassical in nature: i) the outer horizon of the trapped region
contracts, and the inner horizon expands, ultimately pushing past
extremality and evaporating the trapped region completely, and ii) an
anti-trapped region forms beneath the trapped one, having a seemingly
longer lifetime than its trapped counterpart, settling to a size
similar to that of the original BH. Similar results are found in
ref.~\cite{Barenboim2024}, in the context of a two-dimensional dilaton
gravity theory with the same semiclassical perturbation terms (given
that the Polyakov approximation stems from a quantisation in two
dimensions), with the major difference in results being that the
lifetime of the anti-trapped region there can seemingly be much
shorter. For the four-dimensional Einstein-Maxwell background of the
present work, the anti-trapped region appears to be long-lived, as far
as our numerical accuracy has allowed us to discern. However, it is
worth noting that it is very likely that, due to blueshift
amplification, this anti-trapped region of the spacetime would be
unstable under additional perturbations.

While the present results are of interest in terms of showing an
example of backreaction from a semiclassical source and full
evaporation dynamics, there are several factors which may stand
between the present results and the dynamics of realistic
BHs. Firstly, the scale separation between the size of the trapped
region and the Planck length used here is, to put it mildly, not close
to being representative of astrophysical BHs. Yet the qualitative
behaviour appears to remain the same as these scales are brought
further apart, the change in the outcome being only the time it takes
for evaporation to occur. Further analysis to ascertain whether this
tendency remains can be performed in the future with improved
numerical schemes.

Secondly, in generic astrophysical scenarios the collapse and BH
formation process would be more intricate than the simple shell model
we presented here, also containing effectively classical perturbations
in its wake, such as the ones which trigger mass inflation. This issue
can be addressed in future works with more realistic classical matter
ingredients, without changing the numerical solvability of the system,
and still leaving room for the simplification of assuming spherical
symmetry.

Thirdly, the applicability of the Polyakov approximation to obtain the
RSET in the vicinity of the inner apparent horizon needs attention. A
quick comparison between our figure~\ref{Tvv_CauchyHorizon} and
figure~1 of ref.~\cite{Zilberman2019} suggests that the
$\expval{T_{vv}}$ flux component may be similar close to extremality
(note that the standard example of $r_i=0.75$ we use here corresponds
to $Q/M\simeq0.99$). Further away from extremality however, the sign
on the flux changes in the 3+1 dimensional calculation, while it does
not in the Polyakov approximation, suggesting that a detailed analysis
and comparison for a wider class of BH geometries with an inner
horizon, and in different regions of their parameter spaces, is
warranted. Additionally, 3+1 dimensional calculations have only been
performed on stationary backgrounds. The accumulation of light rays at
the inner horizon, in contrast to the peeling away from the outer
horizon, makes it likely that the behaviour of the RSET very close to
the inner horizon becomes increasingly sensitive to the past evolution
of a large region of the spacetime. Dynamical inner horizons can
therefore potentially lead to significant changes in these results,
making the final verdict on the applicability of any approximation
harder to determine.

Lastly, it is important to recall that the the type of field
considered here is a massless neutral scalar, while the background is
a charged BH. Realistic perturbations on such a BH would, of course,
be comprised of both neutral and charged fields, which could lead to
significant differences in terms of their
RSETs~\cite{Montagnon2025}. In a broader sense, it would be worth
exploring the dynamics of systems in which backreaction from the
(approximate) RSET can have an effect on the background inner horizon
scale, such as a charged field on a Reissner-Nordström background, or
any field on the Kerr background (since they all can carry angular
momentum). A similar setup would even be possible with the Polyakov
approximation for a neutral field, if the spherical BH model had a
mass-dependant inner horizon scale.

There are many avenues left to explore in order to deepen our
understanding of the dynamics of BHs with inner horizons. Most
importantly, the range of applicability of the semiclassical theory,
and its dynamical outcome within that range, must be determined before
any full ``quantum gravity" corrections of BHs can be argued to be of
physical interest.

\section*{Acknowledgements}

The authors thankfully acknowledge the computer resources, technical
expertise and assistance provided by CENTRA/IST. Computations were
performed at the cluster “Baltasar-Sete-Sóis” and supported by the
H2020 ERC Advanced Grant “Black holes: gravitational engines of
discovery” grant agreement no. Gravitas–101052587. The work was
furthermore partially supported by PeX-FCT (Portugal) program
2023.12549.PEX and FCT (Portugal) projects UIDB/00099/2020 and
UIDP/00099/2020. VB also acknowledges support from the European
Union’s H2020 ERC Advanced Grant “Black holes: gravitational engines
of discovery” grant agreement no. Gravitas–101052587, as well as from
the Spanish Government through the Grants No. PID2020-118159GB-C43,
PID2020-118159GB-C44, PID2023-149018NB-C43 and PID2023-149018NB-C44
(funded by MCIN/AEI/10.13039/501100011033).

\appendix

\section{Code overview}
\label{AppA}

\begin{figure}[t!]
	\centering
	\includegraphics[scale=0.3]{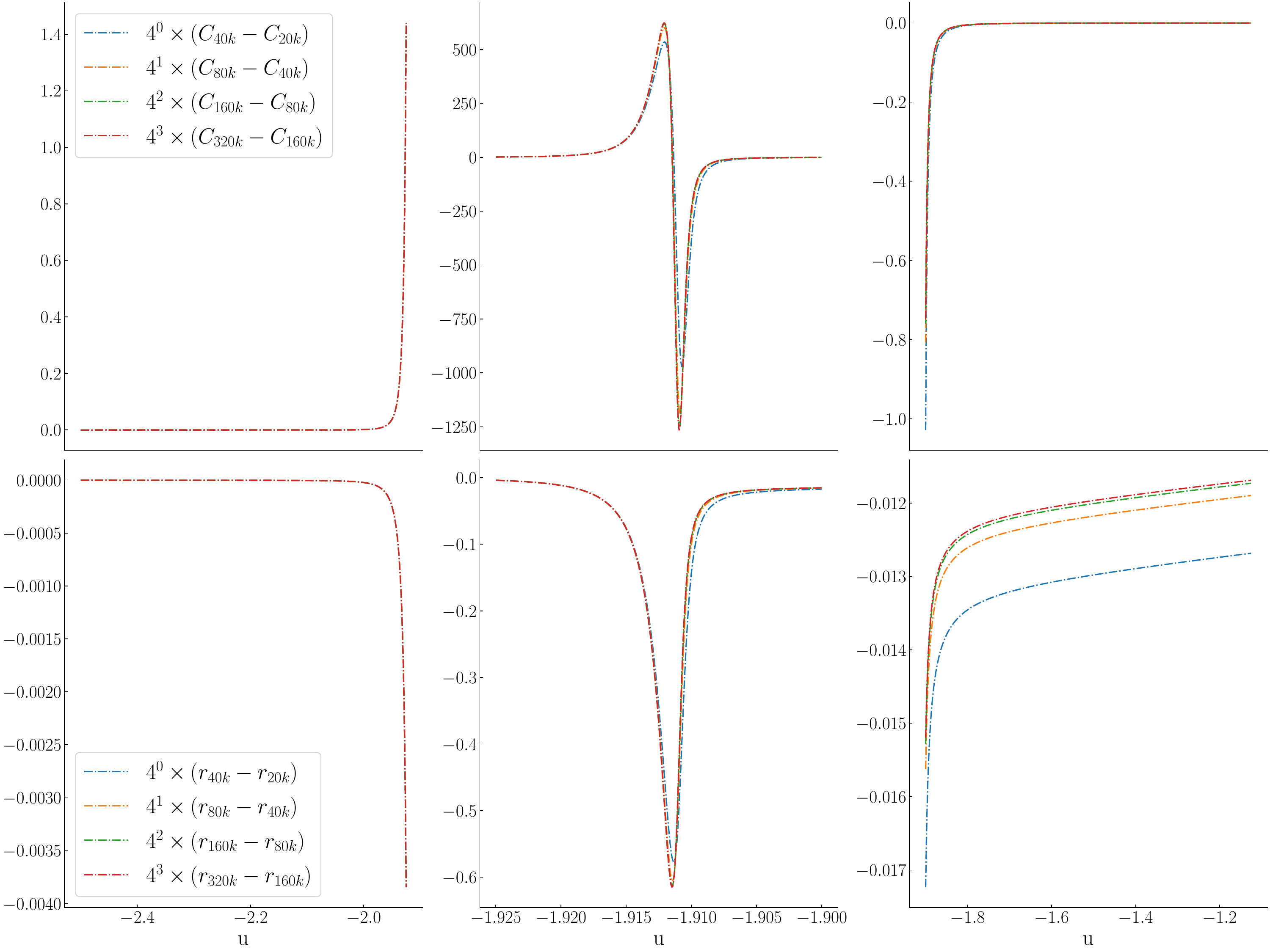}
	\caption{Second order convergence plots for the metric
          functions $r(u,v)$ and $C(u,v)$ with
          $L_p = 0.26, r_- = 0.75$ and $r_+ = 1.0$. The convergence is
          evaluated for a constant advanced time slice $v = 120$ for
          the whole range of $u$. For each variable, the error
          difference is splited in three intervals of $u$ due to the
          high accumulation in error around the anti-trapped surface
          horizon which, for $v = 120$, is situated at $u = -1.896$.}
	\label{Appendix_Convergence_Plot}
\end{figure}

As mentioned in the main text, the evolution code, which is written in
Julia, solves equations \eqref{sev} and \eqref{aev} for the
variables~$s,t,A,B$. (By equality of mixed partials these give four
equations). Thus the variables~$t,B$ are solved for by integrating in
the~$u$-direction, and~$s,A$ in the~$v$-direction. For the metric
components~$r,C$, we simply solve the definitions of the reduction
variables~$t,B$, equations \eqref{rev}, \eqref{cev}, again by
integration in the~$v$-direction. Because of these choices,
\eqref{constraintu}, \eqref{constraintv}, \eqref{sdef} and
\eqref{Adef} are treated as constraints. We thus adopt the standard
free-evolution approach used in numerical relativity. The constraints
are solved for in the initial slices by integration in the appropriate
direction and then monitored throughout the evolution. We also use the
constraints for the purposes of convergence tests.

Since we have not gone to the effort of rewriting the equations in the
nested form common to numerical work in single-null coordinates, the
equations of motion for each direction are coupled, and so we solve
them in tandem. See \cite{Gundlach:2024mld} for a recent discussion of
standard single-null configurations and \cite{Luna:2019olw} for an
example of numerical work in a gauge similar to that used here.

Numerically, we use Heun's method, a simple second order Runge-Kutta
integrator, which has the advantage of not requiring evaluation at
half-steps, and thus avoiding interpolation. Extending the method to
higher-order may be necessary in the future, but would need a little
care because of this subtlety.

We have performed successful convergence tests on precisely the
initial data configurations discussed in the maint text. A
representative example is given in
Fig.~\ref{Appendix_Convergence_Plot}.

\nocite{*}
\bibliography{Bibliografia}
\bibliographystyle{ieeetr}
\end{document}